\documentclass[12pt]{iopart}
\usepackage{graphicx}
\usepackage{braket}
\usepackage{caption}
\usepackage{subfigure}
\usepackage{cancel}
\usepackage{iopams}  
\begin{document}

\title{Entanglement of two movable mirrors with a single photon superposition state}
\footnote{We extend our heartiest congratulations to Margarita and Vladimir Man'ko of a magnificent 150 years and dedicate this article to them.}

\author{Wenchao Ge \& M. Suhail Zubairy}
\address{Institute for Quantum Science and Engineering (IQSE) and Department of Physics \& Astronomy, Texas A\&M University, College Station, Texas 77843, USA}
\vspace{10pt}
\begin{indented}
\item[] 15 December 2014
\end{indented}


\begin{abstract}
We propose a simple scheme to generate deterministic entanglement between two movable end mirrors in a Fabry-Perot cavity using a single photon superposition state. We derive analytically the expressions of the generated entangled states and the degree of entanglement for each state. We show that strong entanglement can be obtained either in the single-photon strong coupling regime deterministically or in the single-photon weak coupling regime conditionally.\end{abstract}

\pacs{03.67 Bg, 42.50 Wk, 42.50 Pq}
%
%
%
%
%

\section{Introduction}

Quantum entanglement is an important phenomenon in quantum physics, which has potential applications in quantum information and quantum computing \cite{Nielsen:10}. Since the birth of quantum mechanics, the idea of entangling a microscopic object and a macroscopic object, the Schrodinger's cat state, drives people to push the limit of quantum mechanics towards the boundary between the quantum world and the classical world ~\cite{Monroe:96,Arndt:99,Gerlich:11,Zurek:03}. Entanglement of microscopic objects has been realized experimentally in the systems of photons \cite{Kwiat:95}, atoms \cite{Hagley:97}, and ions \cite{Turchette:98}. Due to fast decoherence of macroscopic objects \cite{Zurek:03}, quantum entanglement of macroscopic objects remains a difficult task. 

Cavity optomechanics \cite{Aspelmeyer:14}, exploring the effect of radiation-pressure coupling between optical and mechanical elements, provides a platform for realizing quantum effects of macroscopic objects. Optomechanical sideband cooling on a mechanical oscillator \cite{Wilson:07,Marquardt:07,Teufel:11,Chan:11}, normal-mode splitting between optical and mechanical modes \cite{Dobrindt:08,Groblacher:09}, optomechanically induced transparency \cite{Agarwal:10,Weis:10,Safavi:11}, and entanglement between a mechanical oscillator and a cavity field \cite{Marshall:03,Vitali:07,Genes:08,Palomaki:13} have been proposed theoretically and demonstrated experimentally recently.

Many theoretical proposals have been put forward to generate entanglement between two mechanical oscillators in cavity optomechanics: either with coherent driving or without coherent driving. In the former case, entanglement of two mechanical oscillators can be generated using nonclassical states \cite{Zhang:03,Pinard:05,Huang:09,Zhou:11}. Entanglement can also be generated with classical driving fields in a cavity with two mechanical oscillators \cite{Mancini:02,Vitali:07b,Hartman:08,Ludwig:10}, or in two remote optomechanical cavities \cite{Vacanti:08,Borkje:11,Joshi:12}. More recently, ground-state cooling as well as optomechanical entanglement has been proposed in a double-resonant cavity via a correlated emission laser \cite{Xiong:05} with classical driving fields \cite{Ge:13,Ge:13b}. By using coherent driving, optomechanical coupling is affected by the driving laser phase noise \cite{Rabl:09}. To avoid this problem, one may consider entangling two mechanical oscillators using a single photon without coherent driving. Entanglement of two mechanical mirrors in a two-cavity optomechanical system through the coupling of a single photon between cavities has been proposed \cite{Liao:14}. 

In this paper, we propose a simple setup for entangling two movable end mirrors in a Fabry-Perot cavity using a single photon state without any coherent driving. A similar scheme has been proposed \cite{Flayac:14} recently to generate heralded phonon Bell states with very weak coherent driving conditioned on detection of a single photon. Our proposal can generate deterministic entangled phonon states after measurement on the cavity photon state. A single-photon strong coupling rate \cite{Rabl:11,Nunnenkamp:11} may be required for our scheme to generate macroscopic entanglement. Two possible entangled states can be generated depending on the final measurement outcomes. The degree of entanglement of the generated states is quantified by logarithmic negativity and we derive an analytical expression of the negativity for each state. Our results show that the more probable of a state to be generated the smaller the negativity of the state it will be. We also discuss the experimental feasibility of our scheme.

\section{The scheme and the Hamiltonian}

\begin{figure}[t]
\centering
\includegraphics[width =0.48\textwidth]{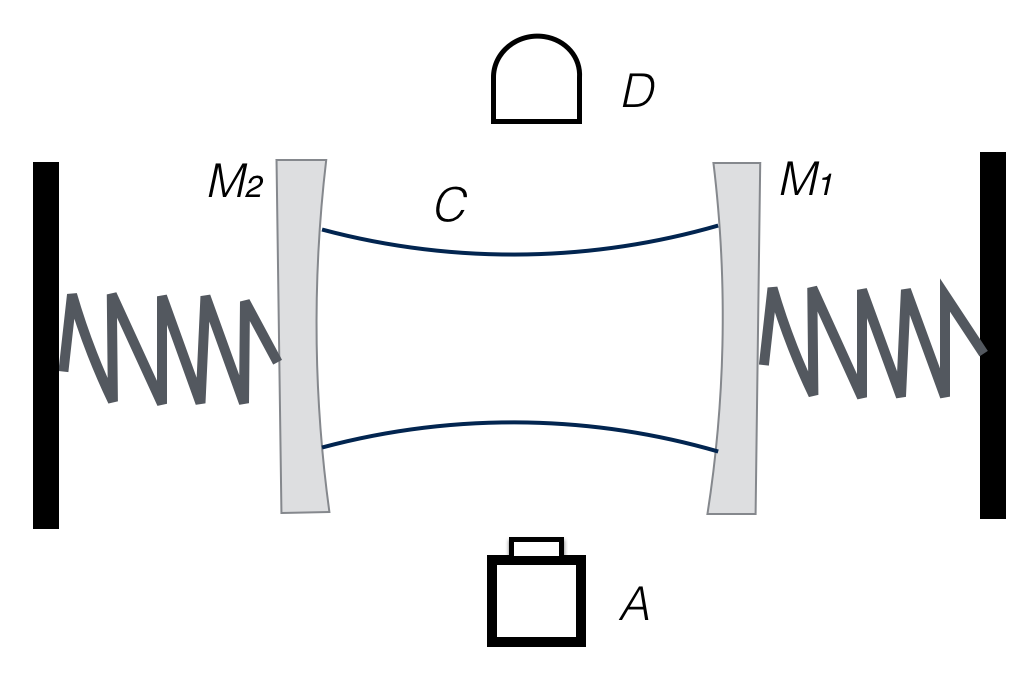}
 \caption{\label{scheme}Experimental setup for entangling two macroscopic mechanical end mirrors $M_1$ and $M_2$. Single-photon superposition state is prepared in the cavity $C$ by sending a superposition atomic state from the single-atom source $A$. After the evolution of the field and the mirrors, another atom in the ground state is sent from $A$ to the cavity. The atomic state is then rotated with a fast $\pi/2$ pulse before detected at the detector $D$. The state of the mirrors then collapses into an entangled state after the detection of the atomic state.}
\end{figure}

We consider a Fabry-Perot cavity with two mechanical end mirrors of masses $m_1, m_2$ and mechanical frequencies $\omega_{m_1}, \omega_{m_2}$ as shown in Fig. \ref{scheme}. These mirrors can be suspended from the ground or attached to cantilevers \cite{Aspelmeyer:14} and we assume only single mode of each mirror is involved in the field-mirror interaction. When light comes into the cavity, it exerts radiation pressure on the mirrors and displaces the mirrors from their equilibrium positions. Therefore, the cavity frequency, which depends on the cavity length $L$, is modulated by the mirrors positions as $\omega_c(\delta x)\approx\omega_c+\frac{\partial \omega_c}{\partial L}\delta x$. Here $\omega_c=n\frac{\pi c}{L}$, $n$ is the integer mode number, and $\delta x = x_1-x_2$ with $x_j$ $(j=1,2)$ the mirror's position. The Hamiltonian of the system is then given by \cite{Law:94}
\begin{eqnarray}
\label{eq:opto}
\mathcal{H} &=& \hbar \omega_c(\delta x) c^{\dagger}c+\sum_{j=1,2}\hbar\omega_{m_j} b_j^{\dagger}b_j\nonumber\\
&=&\hbar \omega_c c^{\dagger}c+\sum_{j=1,2}\hbar\omega_{m_j} b_j^{\dagger}b_j+\hbar\frac{\omega_c}{L} c^{\dagger}c(x_2-x_1),
\end{eqnarray}
where $b_j$ and $c$ are annihilation operators of the $j$'s mirror and the cavity mode, respectively, $x_j=x_{0_j}(b_j^{\dagger}+b_j)$ and $x_{0_j}=\sqrt{\hbar/2m_j\omega_{m_j}}$ is the zero-point fluctuation of the $j$'s mirror. This scheme is simpler than that in Ref. \cite{Liao:14} where two coupled cavities are considered.

The unitary evolution operator of a single mode electromagnetic wave coupled to a mechanical object can be obtained exactly from Baker-Campbell-Hausdorff formula \cite{Bose:97}. Similarly, it is straightforward to derive the unitary evolution operator of our system consisting of one cavity field coupled to two mechanical modes individually. The unitary operator of the system is then given by
\begin{eqnarray}
U(t)&=&e^{-i\omega_c t c^{\dagger}c}e^{i\phi_1(t)(c^{\dagger}c)^2}\mathcal{D}_1\left(\eta_1(t)c^{\dagger}c\right)e^{-i\omega_{m_1} t b_1^{\dagger}b_1}\nonumber\\
&&\times e^{i\phi_2(t)(c^{\dagger}c)^2}\mathcal{D}_2\left(-\eta_2(t)c^{\dagger}c\right)e^{-i\omega_{m_2} t b_2^{\dagger}b_2},
\end{eqnarray}
where $\phi_j(t)=\beta_j^2\left(\omega_{m_j} t-\sin(\omega_{m_j} t)\right)$, the displacement operator for the $j$'s mirror $\mathcal{D}_j(\alpha) = e^{\alpha b_j^{\dagger}-\alpha^{\ast}b_j}$, and $\eta_j(t) = \beta_j(1-e^{-i\omega_{m_j} t})$. Here $\beta_j=\frac{x_{0_j}\omega_c}{L\omega_{m_j}}$ which characterizes the ratio of the displacement of a single photon due to radiation pressure comparing to that of the mirror's zero-point fluctuation. The terms proportional to $(c^{\dagger}c)^2$ are effective Kerr-like terms of the cavity in the presence of mechanical mirrors. In the single-photon strong coupling regime when $\beta_j\gtrsim1$, photon blockade effect can happen due to the Kerr-like terms \cite{Rabl:11,Nunnenkamp:11}. The argument  $\eta_j(t)c^{\dagger}c$ in the displacement operator means that the displacement on each mirror is proportional to the photon number $c^{\dagger}c$ if the cavity field is initially prepared in a Fock state. Therefore, a superposition state of the cavity photon number can result in a superposition of displacement of the mechanical mirror \cite{Bose:97}.

\section{Entanglement generation}

\subsection{System evolution}
We consider initially the mirrors are cooled to their ground states $\ket{0}_{m_j}$ via optomechanical cooling. A superposition state $\ket{\psi_c}=\frac{1}{\sqrt{2}}(\ket{0}_c+\ket{1}_c)$ is prepared inside the cavity. Due to the radiation pressure, the cavity photon state evolves together with the mechanical mirrors. The state of the system is given by
\begin{eqnarray}
\ket{\psi_s(t)}&=&U(t)\ket{\psi_c}\ket{0}_{m_1}\ket{0}_{m_2} \nonumber\\
&=&\frac{1}{\sqrt{2}}\ket{0}_c\ket{0}_{m_1}\ket{0}_{m_2}+\frac{1}{\sqrt{2}}e^{i2\phi(t)}\ket{1}_c\ket{\eta(t)}_{m_1}\ket{-\eta(t)}_{m_2},
\end{eqnarray}
where the mirrors' states $\ket{\eta(t)}_{m_1}$ and $\ket{-\eta(t)}_{m_2}$ are coherent states due to the radiation pressure of the single photon. Here we consider, for simplicity, $m_1=m_2=m$, $\omega_{m_1}=\omega_{m_2}=\omega_m$, $\phi_1(t)=\phi_2(t)=\phi(t)$ and $\eta_1(t)=\eta_2(t)=\eta(t)$. From the expression of $\ket{\psi_s(t)}$, the system of one cavity field and two-mechanical mirrors are entangled in general. 

After integer number of the mirrors' oscillations $\tau_n=2n\pi/\omega_m$, the system evolves to its initial state, namely $\ket{\psi_s (\tau_n)}=\ket{\psi_s (0)}$. There is no entanglement among the field and the mirrors. After half integer number of the mirrors' oscillation $\tau_{n+\frac{1}{2}}=(2n+1)\pi/\omega_m$, the displacement of the mirrors are largest where $\ket{\psi_s (\tau_{n+\frac{1}{2}})}=\frac{1}{\sqrt{2}}(\ket{0}_c\ket{0}_{m_1}\ket{0}_{m_2}+e^{i\theta_n}\ket{1}_c\ket{2\beta}_{m_1}\ket{-2\beta}_{m_2})$ with $\beta=\beta_1=\beta_2$ and $\theta_n=2\pi(2n+1) \beta^2$. In this case, the entanglement among the tripartite is the maximum. 

To see whether there is entanglement between the mirrors due to radiation pressure only, we trace over the cavity field state. We obtain the mirrors' state as 
\begin{eqnarray}
\rho_m(t)&=&\Tr_c\braket{\ket{\psi_s(t)}\bra{\psi_s(t)}}\nonumber\\
              &=&\frac{1}{2}\left(\ket{0}_{m_1}\ket{0}_{m_2}\bra{0}_{m_2}\bra{0}_{m_1}\right.\nonumber\\
              &+&\left.\ket{\eta(t)}_{m_1}\ket{-\eta(t)}_{m_2}\bra{-\eta(t)}_{m_2}\bra{\eta(t)}_{m_1}\right),
\end{eqnarray}
which is a mixed and separable state.
 
\subsection{Entanglement via measurement}
Generation of entanglement of quantum objects, such as atoms \cite{Plenio:99} and superconducting qubits \cite{Li:08}, have been considered by continuously monitoring the cavity field that interacts with these quantum objects. If no photon is detected outside the cavity, the system of the quantum objects is prepared in an entangled state. Here we show that by measuring the state of the field inside the cavity, the mirrors collapse to an entangled state. Depending on the measurement result, there are two possible entangled states of the mirrors. We study the properties of each state in the following.

To measure the photon number state of the cavity, we first map the cavity field state to a flying two-level atom by interacting with each other for a $\pi$ Rabi oscillation $U_{\pi}$. Then we apply a fast $\pi/2$ pulse $R_{\frac{\pi}{2}}$ on the atomic state such that $R_{\frac{\pi}{2}}\ket{g}_a=(\ket{g}_a+\ket{e}_a)/\sqrt{2}$ and $R_{\frac{\pi}{2}}\ket{e}_a=(-\ket{g}_a+\ket{e}_a)/\sqrt{2}$, where $\ket{g}_a$ ($\ket{e}_a$) is the ground (excited) state of the atom. This process is assumed to be much faster than the mirrors' oscillation frequency $\omega_m$ and the interaction time can be neglected. The system becomes
\begin{eqnarray}
R_{\frac{\pi}{2}}U_{\pi}\ket{\psi_{s}(\tau_{n+\frac{1}{2}})}\ket{g}_a=\left(\sqrt{\mathcal{P}_{-}}\ket{g}_a\ket{\psi_{m-}}+\sqrt{\mathcal{P}_{+}}\ket{e}_a\ket{\psi_{m+}}\right)\ket{0}_c,
\end{eqnarray}
 where 
 \begin{equation}
 \label{eq:entangle}
 \ket{\psi_{m\pm}}=\frac{1}{\sqrt{4\mathcal{P}_{\pm}}}\left(\ket{0}_{m_1}\ket{0}_{m_2}\pm e^{i\theta_n}\ket{2\beta}_{m_1}\ket{-2\beta}_{m_2}\right),
 \end{equation}
 and $\mathcal{P}_{\pm}=\frac{1}{2}\pm\frac{1}{2}\cos(\theta_n)e^{-4\beta^2}$. The state of the mirrors then collapse to the state $\ket{\psi_{m_{+}}}$ ($\ket{\psi_{m_{-}}}$) with probability $\mathcal{P}_{+}$ ($\mathcal{P}_{-}$) after we make a measurement on the atomic state and find the atom in the state $\ket{e}_a$ ($\ket{g}_a$).  After the measurement the mirrors are disentangled with the atom and the cavity is in vacuum,  therefore the entangled state of the mirrors is under free evolution. 

For a single-photon strong coupling rate, i. e., $\beta\gtrsim1$, ${}_{m_1}\langle 0 | 2\beta\rangle_{m_1}= \mathstrut_{m_2}\langle 0|-2\beta\rangle_{m_2}=e^{-2\beta^2}\ll1$. The two parts of the entangled states, $\ket{0}_{m_1}\ket{0}_{m_2}$ and $\ket{2\beta}_{m_1}\ket{-2\beta}_{m_2}$, are almost orthogonal to each other. Therefore, the entanglement is strong in this regime.

For a single-photon weak coupling rate, i. e., $\beta\ll1$,  
$\ket{\psi_{m+}}\approx \ket{0}_{m_1}\ket{0}_{m_2}+\beta(\ket{1}_{m_1}\ket{0}_{m_2}-\ket{0}_{m_1}\ket{1}_{m_2})$ with $\mathcal{P}_{+}\approx 1-2\beta^2$ while $\ket{\psi_{m-}}\approx \frac{1}{\sqrt{2}}(\ket{1}_{m_1}\ket{0}_{m_2}-\ket{0}_{m_1}\ket{1}_{m_2})$ with $\mathcal{P}_{-}\approx 2\beta^2$. The two states $\ket{\psi_{m_{\pm}}}$ are almost orthogonal to each other. Although $\mathcal{P}_{-}\ll1$, the state $\ket{\psi_{m-}}$ is a phonon Bell state \cite{Flayac:14} which is a strong entanglement state. Therefore, to realize the state $\ket{\psi_{m-}}$ in the weak single-photon coupling regime, one needs to repeat the experiment many times to get one measurement result of $\ket{g}_a$.

\begin{figure*}
\centering
 \subfigure[]{
\includegraphics[width =0.45\linewidth]{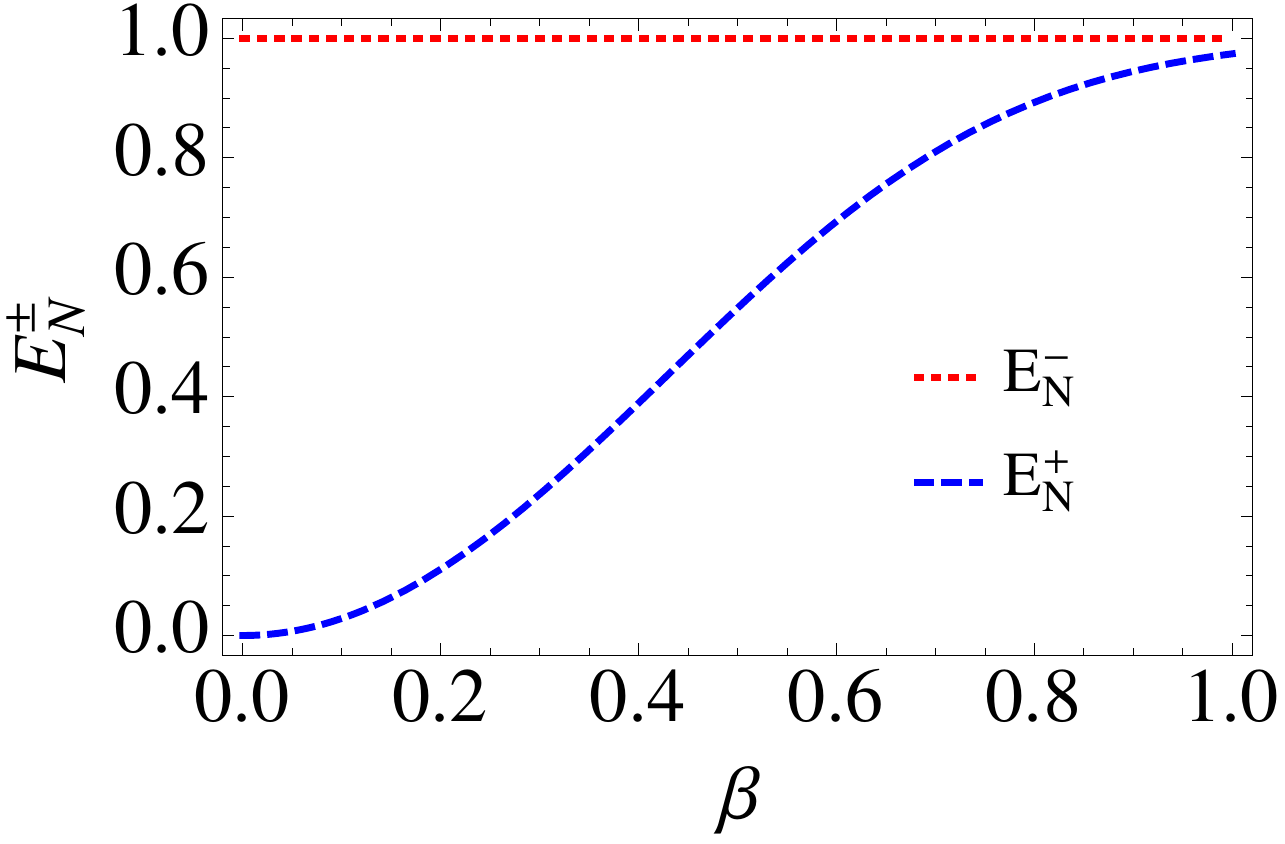}
\label{fig:entanglement1}
 }
 \subfigure[]{
\includegraphics[width =0.45\linewidth]{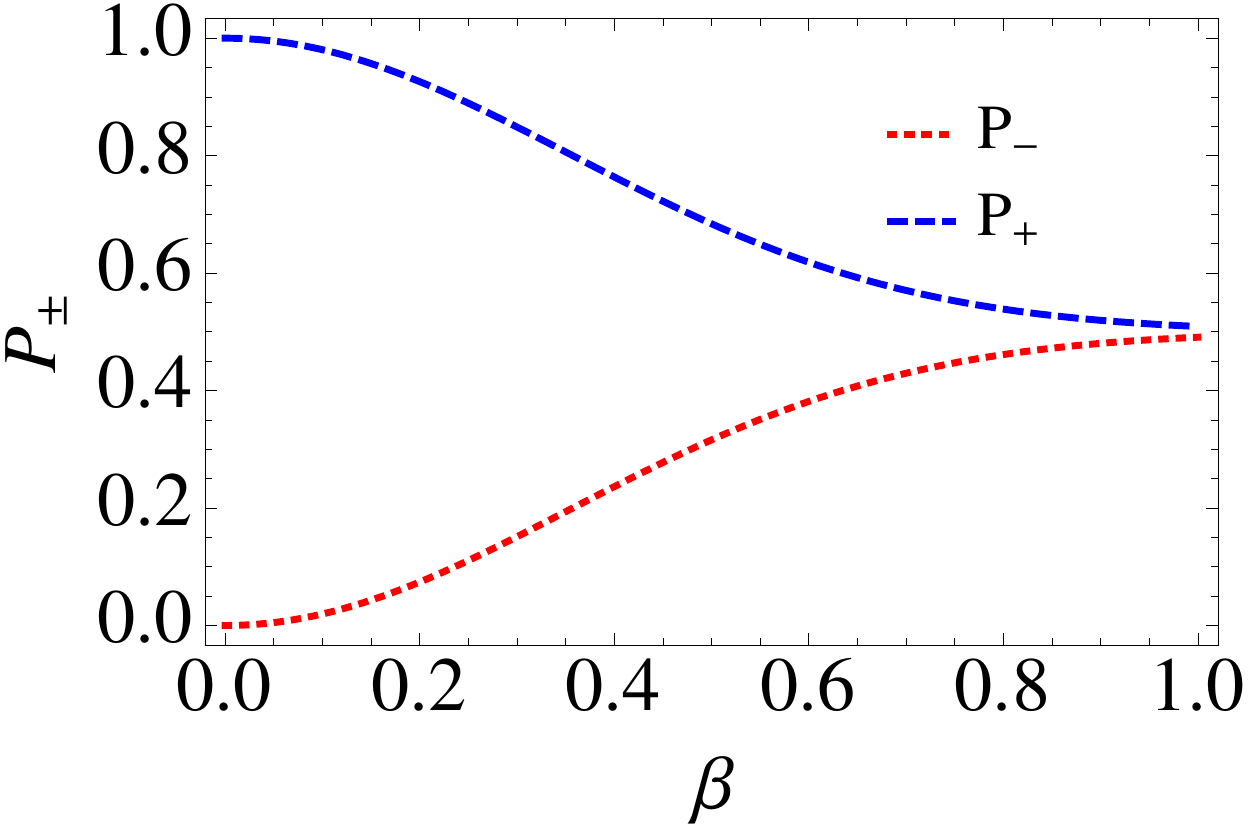}
\label{fig:probability1}
 }
\caption{(color online).(a) Logarithmic negativity $E_{\mathcal{N}}^{\pm}$ and (b) the corresponding probability $\mathcal{P}_{\pm}$ of entangled states $\ket{\psi_{m\pm}}$ versus the single-photon coupling rate $\beta$ for zero phase.}
\label{fig:zero}
\end{figure*}

\subsection{Entanglement quantification}

\subsubsection{Entanglement measure}
We have shown for $\beta\gtrsim1$, the entanglement for both states $\ket{\psi_{m_{\pm}}}$ is strong since the two parts in each state are nearly orthogonal. For $\beta\ll1$, we have a low probability of $2\beta^2$ to obtain a strong entanglement.  
Now we quantify the degree of entanglement of our system in general using the logarithmic negativity \cite{Vidal:02}.  A quantum state $\rho$ of a bipartite system with two subsystems $A$ and $B$ is defined to be separable if and only if $\rho$ can be
written as
\begin{eqnarray}
\rho=\sum_{j}p_j\rho_{jA}\otimes\rho_{jB},
\end{eqnarray}
with $\sum_j p_j=1$ and $0\le p_j\le1$. A number of entanglement measures \cite{Peres:96,Horodecki;96,Duan:00,Simon:00,Vidal:02,Hillery:06,Sun:09} have been proposed based on the physicality of the partial transposed density operator $\rho^{T_A}$. The logarithmic negativity for a continuous variable bipartite system is defined as \cite{Vidal:02}
\begin{eqnarray}
E_{\mathcal{N}}(\rho)\equiv \mathrm{log_2}||\rho^{T_A}||_1,
\end{eqnarray}
where $||\rho^{T_A}||_1$ is the trace norm of the partial transposed density operator $\rho^{T_A}$. The general expression of logarithmic negativity is cumbersome to calculate since the density matrix has an infinite dimension as can be seen from Eq. (\ref{eq:entangle}). However, for any pure state $\rho=\ket{\Phi}\bra{\Phi}$, the logarithmic negativity can be readily calculated as 
\begin{eqnarray}
E_{\mathcal{N}}(\rho)= \mathrm{log_2}\Big(\sum_k c_k\Big)^2,
\end{eqnarray}
where $c_k$ are coefficients of the Schmidt decomposed state $\ket{\Phi}=\sum_k c_k\ket{e_k}_A\otimes\ket{e_k}_B$. Here $\ket{e_k}_A$ and $\ket{e_k}_B$ are orthonormal basis of the two subsystems after the Schmidt decomposition. As will be shown in the following, the state of Eq. (\ref{eq:entangle}) can be Schmidt decomposed into a state of two dimensions whose logarithmic negativity is straightforward to calculate.

\subsubsection{Zero phase}
We consider, for simplicity, that $e^{i\theta_n}\approx1$ for both strong and weak single-photon coupling regimes, therefore we neglect this phase term in the expression of $\ket{\psi_{m_{\pm}}}$. This condition can be satisfied by varying the number of oscillations $n$ for different values of $\beta$. First, we decompose the state $\ket{\psi_{m_-}}$ as
\begin{equation}
\ket{\psi_{m-}}=\frac{1}{\sqrt{2}}\sum\limits_{k=1,2}\ket{A_k}_{m_1}\ket{A_k}_{m_2},
\end{equation}
where $\ket{A_1}_{m_j}=(-1)^j(c_1^{-}\ket{0}+c_2^{-} \widetilde{\ket{2\beta}})_{m_j}$, $\ket{A_2}_{m_j}=(c_2^{-}\ket{0}-c_{1}^{-}\widetilde{\ket{2\beta}})_{m_j}$, $c_k^{-}=\sqrt{\big(1+(-1)^k\sqrt{1-e^{-4\beta^2}}\big)/2}$, and $\widetilde{\ket{2\beta}}_{m_j}=\frac{1}{\sqrt{1-e^{-4\beta^2}}}(\ket{(-1)^{j+1}2\beta}_{m_j}-e^{-2\beta^2}\ket{0}_{m_j})$. The negativity of the state  $\ket{\psi_{m_-}}$ is $E_{\mathcal{N}}^{-}=1$ which equals to the maximum value of the logarithmic negativity of $2\times2$ system, i. e. the logarithmic negativity of a Bell state. The property of strong entanglement for this state holds for all values of $\beta$ even if $\beta\ll1$, albeit it happens with a small probability for small $\beta$. $\ket{\psi_{m_+}}$ can be decomposed as
\begin{equation}
\ket{\psi_{m_+}}=\frac{1}{\sqrt{4\mathcal{P}_{+}}}\sum\limits_{k=1,2}c^{+}_k\ket{B_k}_{m_1}\ket{B_k}_{m_2},
\end{equation}
where $c^{+}_{j}=1+(-1)^je^{-2\beta^2}$, $\ket{B_1}_{m_j}=(\sqrt{c^{+}_1/2}\ket{0}-\sqrt{c^{+}_2/2} \widetilde{\ket{2\beta}})_{m_j}$, and $\ket{B_2}_{m_j}=(\sqrt{c^{+}_2/2}\ket{0}+\sqrt{c^{+}_1/2} \widetilde{\ket{2\beta}})_{m_j}$. The negativity of this state is then given by $E_{\mathcal{N}}^{+}=1-\log_2(1+e^{-4\beta^2})$, which shows weak entanglement for $\beta\ll1$ and strong entanglement for $\beta\gg1$. We plot the results $E_{\mathcal{N}}^{\pm}$ of the two states in Fig. \ref{fig:entanglement1} and the probabilities $\mathcal{P}_{\pm}$ to obtain these states in Fig. \ref{fig:probability1}. We observe from both figures that $E_{\mathcal{N}}^{-}>E_{\mathcal{N}}^{+}$ while the relation of the probabilities $\mathcal{P}_{\pm}$ is the inverse.

\subsubsection{Non-zero phase}
\begin{figure*}
\centering
 \subfigure[]{
\includegraphics[width =0.45\linewidth]{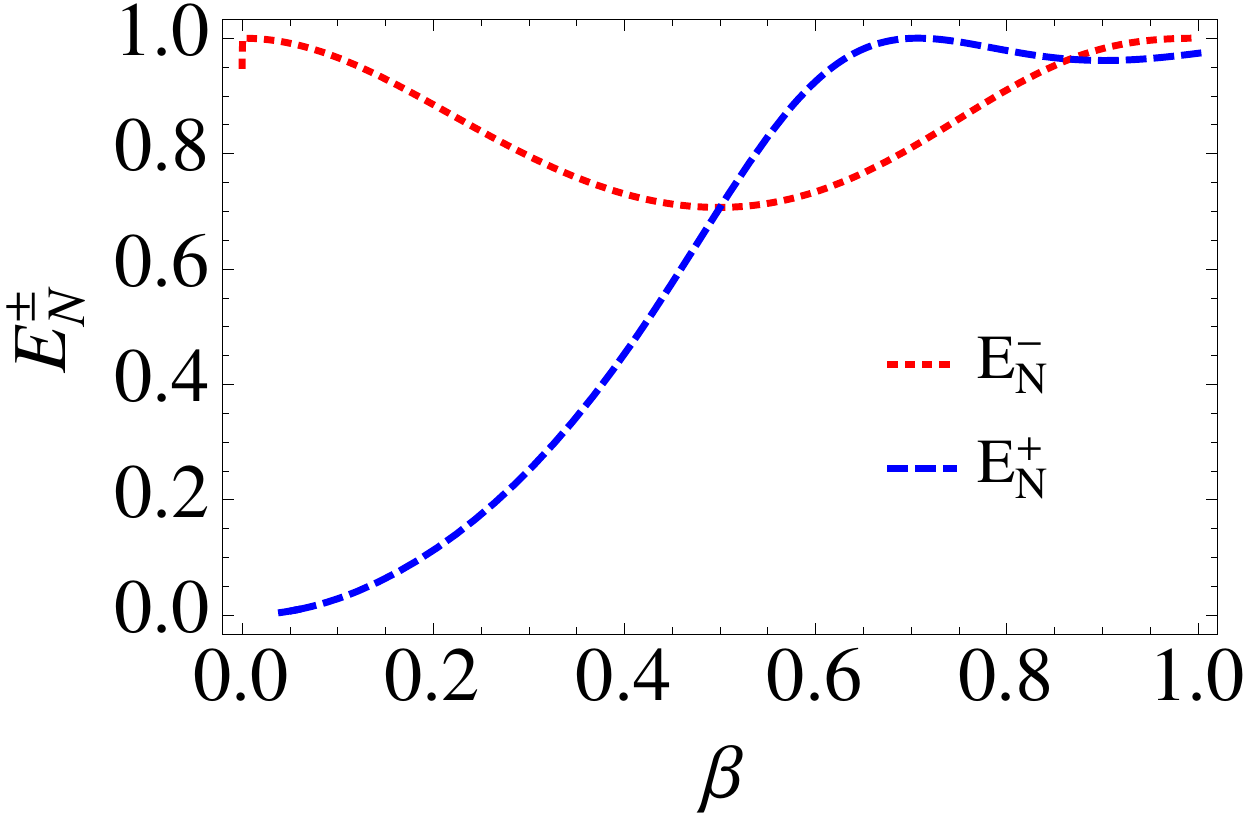}
\label{fig:entanglement2}
 }
 \subfigure[]{
\includegraphics[width =0.45\linewidth]{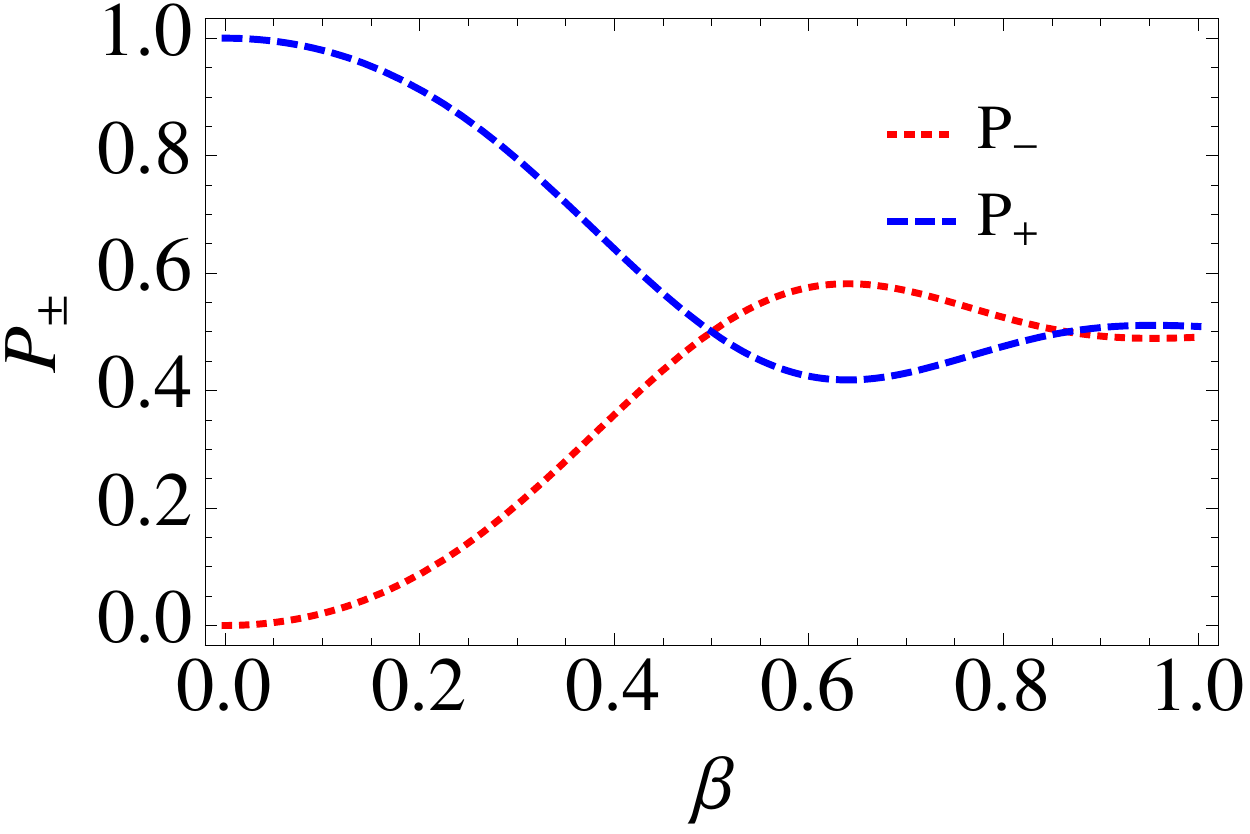}
\label{fig:probability2}
 }
\caption{(color online).(a) Logarithmic negativity $E_{\mathcal{N}}^{\pm}$ and (b) the corresponding probability $\mathcal{P}_{\pm}$ of entangled states $\ket{\psi_{m\pm}}$ versus the single-photon coupling rate $\beta$ including the phase term $e^{i\theta_0}$.}
\label{fig:arbitrary}
\end{figure*}

Now we calculate the general case of the logarithmic negativity of the two states $\ket{\psi_{m\pm}}$ including the phase term $e^{i\theta_n}$. The generated states $\ket{\psi_{m\pm}}$ can be decomposed into $\sum\limits_{k=1,2}d_k^{\pm}\ket{C_{k}^{\pm}}_{m_1}\ket{D_{k}^{\pm}}_{m_2}$, where $\ket{C_{k}^{\pm}}_{m_1}$ and $\ket{D_{k}^{\pm}}_{m_2}$ are the mirrors' states under Schmidt decomposition and 
\begin{equation}
d_k^{\pm}=\sqrt{\frac{1}{2}+(-1)^k\frac{1}{2}\sqrt{1-\Big[\frac{1-e^{-4\beta^2}}{1\pm e^{-4\beta^2}\cos(\theta_n)}\Big]^2}}.
\end{equation}
We do not provide the explicit expressions of $\ket{C_{k}^{\pm}}_{m_1}$ and $\ket{D_{k}^{\pm}}_{m_2}$ since they are not important in calculating the negativity. The logarithmic negativity is the given by $E_{\mathcal{N}}^{\pm}=2\log_2(d_1^{\pm}+d_2^{\pm})$. We plot the degree of entanglement and the corresponding probability in this case for $\theta_0=2\pi \beta^2$ in Fig. \ref{fig:arbitrary}. We observe from the figure that the degree of entanglement as well as the probability oscillates back and forth with increasing single-photon coupling rate due to the $\cos(\theta_n)$ term. It is not difficult to see from the expression of $\cos(\theta_n)$  that for greater number $n$ these curves will have more frequent oscillations. By comparing the results of the two states  $\ket{\psi_{m\pm}}$, we also observe that the greater the negativity the smaller the probability.

\subsection{Finite temperature case}
For a finite temperature $T$, the mirrors are initially prepared in thermal states given by $\rho_{m}^0(n_{th})=\frac{1}{(\pi n_{th})^2}\int d^2\alpha_1 e^{-|\alpha_1|^2/n_{th}} (\ket{\alpha_1}\bra{\alpha_1})_{m_1}\otimes\int d^2\alpha_2 e^{-|\alpha_2|^2/n_{th}} (\ket{\alpha_2}\bra{\alpha_2})_{m_2}$ with thermal phonon number $n_{th}=(e^{\hbar\omega_m/k_B T}-1)^{-1}$ for both mirrors and $k_B$ the Boltzmann constant. After the interaction steps described above, the final state of the mirror will be given by the density operator 
\begin{eqnarray}
\rho_{m\pm}(n_{th})&=&\frac{1}{(\pi n_{th})^2}\int \int d^2\alpha_1d^2\alpha_2  e^{-(|\alpha_1|^2+|\alpha_2|^2)/n_{th}}\nonumber\\
&\times&\ket{\psi_{m\pm}(\alpha_1,\alpha_2)}\bra{\psi_{m\pm}(\alpha_1,\alpha_2)},
\end{eqnarray}
 where
 \begin{eqnarray}
\ket{\psi_{m\pm}(\alpha_1,\alpha_2)}=\frac{1}{\sqrt{4\mathcal{P}_{\pm}}}\big(\ket{\alpha_1}_{m_1}\ket{\alpha_2}_{m_2}\pm e^{i\theta_n}\ket{\alpha_1+2\beta}_{m_1}\ket{\alpha_2-2\beta}_{m_2}\big).
\end{eqnarray}
The final density operator $\rho_{m\pm}(n_{th})$ represents a mixed state of entangled states. The degree of entanglement of such a state is very complicated to calculate and we do not discuss it here.

\section{Discussion and conclusion}
In this paper, we proposed a simple setup to generate deterministic entanglement between two movable end mirrors in a Fabry-Perot cavity using a single photon state. We discussed two single-photon coupling regimes of entanglement generation. In the weak single-photon coupling regime, strong entanglement can be generated with a very small probability. In the strong single-photon coupling regime, strong entanglement can always be generated. Our scheme can be extended to superconducting circuits \cite{Pirkkalainen:13} and nano-resonators coupled to nitrogen-vacancy centers \cite{Kolkowitz:12} where single-photon strong coupling rate is possible. Another possibility of generating strong entanglement without a strong single-photon coupling rate is to magnify the coherent states of the mirrors via periodic qubit flipping \cite{Asadian:14,Ge:15}.  By comparing the two possible outcomes of the entangled states, we also observed that the smaller the probability to generate an entangled state the stronger the degree of entanglement of that state.

We now discuss the experimental realization of our scheme. Single-photon superposition state can be generated using a single atom in a superposition state interacting for a period of $\pi$ Rabi rotation \cite{Raimond:01}. We require the atom-field coupling strength $g_c\gg\omega_m$ such that the mirrors' motion does not affect the atom-field interaction. This also guarantees the fast mapping between the photonic state and the atomic state during the measurement process. We also assume the decay rate of the photon $\kappa\ll\omega_m$ in order to keep the photon inside the cavity during the interaction. Ground-state cooling of the mirrors has been demonstrated in optomechanics recently \cite{Teufel:11,Chan:11} in this regime. In summary, our scheme of generating macroscopic entangled states may be realized in experiment.

\section*{ACKNOWLEDGEMENT}
This research is supported by an NPRP Grant (No. 5-102-1-026) from Qatar National Research Fund.

\section*{References}
\bibliography{Macroscopic_Optomechanics_from_single-photon_measurements}

\end{document}